# Strain Engineering of Magnetoresistance and Magnetic Anisotropy in CrSBr

Eudomar Henríquez-Guerra[1,#], Alberto M. Ruiz[2,#], Marta Galbiati[2], Alvaro Cortes-Flores[1], Daniel Brown[2], Esteban Zamora-Amo[3], Lisa Almonte[1], Andrei Shumilin[2], Juan Salvador-Sánchez[4], Ana Pérez-Rodríguez[4], Iñaki Orue[5], Andrés Cantarero[2], Andres Castellanos-Gomez[3], Federico Mompeán[3], Mar Garcia-Hernandez[3], Efrén Navarro-Moratalla[2], Enrique Díez[4], Mario Amado[4,6], José J. Baldoví[2,*], and M. Reyes Calvo[1,7,*]

[1] BCMaterials, Basque Center for Materials, Applications and Nanostructures, 48940 Leioa, Spain

[2] Instituto de Ciencia Molecular, Universitat de València, 46980 Paterna, Spain

[3] 2D Foundry Research Group, Instituto de Ciencia de Materiales de Madrid (ICMM-CSIC), Madrid E-28049, Spain

[4] Nanotechnology Group, USAL–Nanolab, Departamento de Física Fundamental, University of Salamanca, 37008 Salamanca, Spain

[5] SGIker Medidas Magnéticas, 48940 Leioa, Spain

[6] IUFFyM, University of Salamanca, 37008, Salamanca, Spain

[7] IKERBASQUE, Basque Foundation for Science, 48009 Bilbao, Spain

[#] equal contribution

* email: j.jaime.baldovi@uv.es, reyes.calvo@bcmaterials.net

## Abstract

Tailoring magnetoresistance and magnetic anisotropy in van der Waals magnetic materials is essential for advancing their integration into technological applications. In this regard, strain engineering has emerged as a powerful and versatile strategy to control magnetism at the two-dimensional (2D) limit. Here, we demonstrate that compressive biaxial strain significantly enhances the magnetoresistance and magnetic anisotropy of few-layer CrSBr flakes. Strain is efficiently transferred to the flakes from the thermal compression of a polymeric substrate upon cooling, as confirmed by temperature-dependent Raman spectroscopy. This strain induces a remarkable increase in the magnetoresistance ratio and in the saturation fields required to align the magnetization of CrSBr along each of its three crystalographic directions, reaching a twofold enhancement along the magnetic easy axis. This enhancement is accompanied by a subtle reduction of the Néel temperature by ~10K. Our experimental results are fully supported by first-principles calculations, which link the observed effects to a strain-driven modification in interlayer exchange coupling and magnetic anisotropy energy. These findings establish strain engineering as a key tool for fine-tuning magnetotransport properties in 2D magnetic semiconductors, paving the way for implementation in spintronics and information storage devices.

## Introduction

Magnetic anisotropy, the energy required to reorient the magnetization of a material, is a key ingredient in modern magnetic devices. It plays a crucial role in magnetotransport properties and underpins phenomena such as magnetoresistance, a cornerstone in the development of high-density magnetic memories. The recent discovery of intrinsic long-range magnetic order in two-dimensional (2D) materials has opened exciting avenues in the exploration of low-dimensional magnetism[1–3]. Notably, the presence of highly anisotropic tunable phases in van der Waals magnets[4–7] enables the realization of diverse magnetic ground states, positioning them as promising platforms for next-generation spintronics, information storage and quantum technologies[8]. Indeed, magnetoresistive switching has been recently demonstrated in atomically thin semiconducting magnets[7,9,10]. Recent breakthroughs include non-volatile electrical control[11] and giant magnetoresistance in 2D magnetic tunnel junctions[12]. In this class of materials, magnetic anisotropy not only governs magnetotransport characteristics but is also essential for stabilizing magnetic order at the few-layer limit[2,13]. The intrinsic sensitivity of 2D materials to external perturbations opens the door to fine-tuning their properties, and thus to achieving precise control over 2D magnetism.

Strain engineering has emerged as a versatile and powerful strategy to modulate the electronic and magnetic properties of 2D materials[14,15]. Theoretical studies predict that strain can tune, enhance or even induce a reversible switching of magnetic order in layered vdW magnets[16–22]. Furthermore, strain has been proposed as an efficient tool to control magnetic anisotropy[20,23–25] and optimize magnon transport[26]. Recent experimental advances have demonstrated tensile strain-induced switching between antiferromagnetic and ferromagnetic states in CrSBr[27–29], mapped its magnetization under non-uniform tensile strain[30], and reported a strain-driven decrease in the Néel temperature ($T_N$) in $FePS_3$ nanodrums[31]. Nevertheless, these studies rely predominantly on optical and local probe methods. The role of strain on device magnetoresistance has only recently been reported, with CrSBr tunnel junctions showing magnetic switching behavior under uniaxial tensile strain[29]. In contrast, direct transport evidence of strain-enhanced magnetic anisotropy remains scarce, and the impact of compressive strain in tuning 2D magnetism is largely unexplored, primarily due to the limited availability of suitable experimental techniques. In that sense, thermal mismatch strain engineering has emerged as a powerful approach for tuning low-temperature phenomena in 2D materials[32]. This method exploits the differential thermal expansion between a crystal and its substrate. Recently, it has been applied to tune low-temperature excitonic properties in single-layer transition metal dichalcogenides[32,33] and to modulate the superconducting phase transition in few-layer $NbSe_2$ flakes[34], offering a promising, yet unexplored, approach for tuning 2D magnetism in van der Waals materials.

Among the family of van der Waals magnetic materials, the layered A-type, antiferromagnetic semiconductor CrSBr—characterized by ferromagnetic intralayer and antiferromagnetic interlayer coupling and a Néel temperature of $T_N$ = 132 K—is attracting a particularly large attention[35]. This material exhibits highly anisotropic electronic and magnetic properties [5,22,36–38]. Furthermore, CrSBr displays a unique interplay between excitons and magnetism[39–41], exceptional air stability[9,35] and giant magnetoresistance[9,42], making it a prime candidate for spintronic applications[11,12]. In this work, we demonstrate the experimental modulation of magnetoresistance and magnetic anisotropy in few-layer CrSBr flakes through biaxial compressive strain engineering. Our experimental findings are supported by first-principles calculations, which reveal how compressive strain fine-tunes exchange coupling interactions and magnetic anisotropy energy, thereby enabling precise control over the magnetic behavior of CrSBr.

## Results

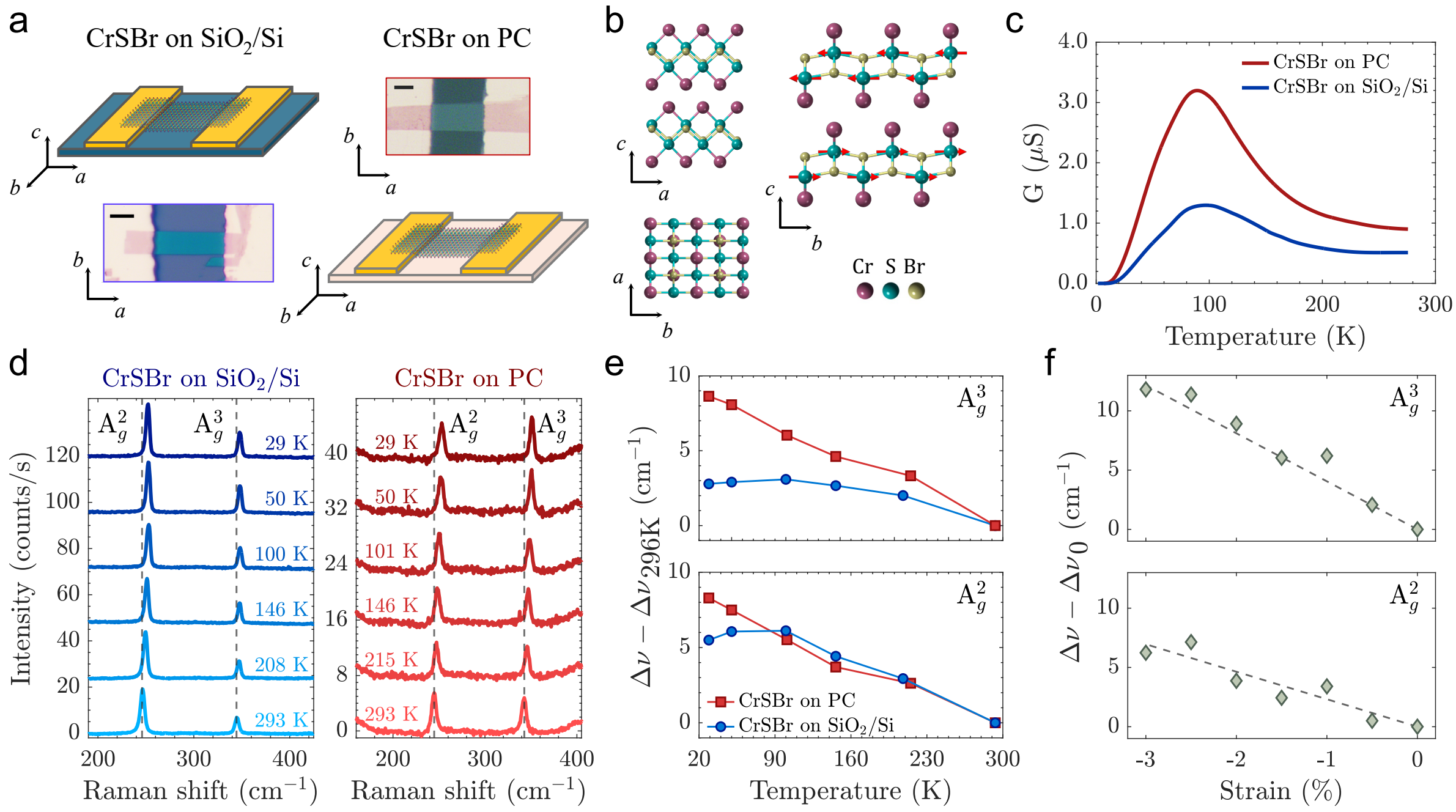

**Figure 1. Conductance and Raman spectroscopy as a function of temperature. a)** Schematic representation and optical microscopy images of CrSBr flakes deposited on pre-patterned Ti/Au electrodes on $SiO_2$/Si (left) and polycarbonate (PC, right) substrates, respectively. Scale bar represents 10 $\mu$m in the optical images. **b)** Crystal structure of CrSBr. Arrows indicate the magnetization orientation along the crystallographic *b*-axis (magnetic easy axis), showing intralayer ferromagnetic and interlayer antiferromagnetic couplings. **c)** Two-terminal conductance (*G*) as a function of temperature for CrSBr on the $SiO_2$/Si and PC devices represented in panel a. **d)** Variable-temperature Raman spectra for CrSBr flakes of similar thickness to those in panel a, deposited on $SiO_2$/Si (left) and PC (right), respectively, acquired under unpolarized laser excitation. A linear background has been subtracted to data for samples on PC. Spectra are vertically shifted for clarity. The $A_g^2$ and $A_g^3$ peak positions at room temperature are marked by dashed gray lines. **e)** Thermal evolution of the $A_g^2$ (bottom) and $A_g^3$ (top) modes, relative to their respective positions at room temperature. **f)** Computed evolution of the $A_g^2$ and $A_g^3$ Raman modes frequency with applied biaxial compressive strain, relative to their respective positions at zero strain, extracted from phonon calculations. The dashed lines represent a linear fit to data.

Thin CrSBr flakes, with thickness of (9 ± 1) nm (see Methods), were prepared via mechanical exfoliation. CrSBr exfoliates naturally into rectangular flakes, with the long edge aligned along the *a* crystallographic axis[43] (Figure 1a-b) and the short edge along the *b*-axis, which corresponds to the easy axis of magnetization (Figure 1b). Using a dry-transfer method, these flakes were deposited onto polycarbonate (PC) substrates with pre-patterned Ti/Au electrodes (Figure 1a). The high thermal expansion coefficient and large Young's modulus of PC make this polymer particularly effective for the transfer of biaxial compressive strain to 2D materials when samples are cooled down to cryogenic temperatures (see Refs. [33,34,44]).

Additionally, we prepared control samples with similarly sized CrSBr flakes, i.e. comparable thickness and in-plane dimensions, onto $SiO_2$/Si substrates. As will be shown later, the simultaneous characterization of CrSBr flakes on both substrates allows us to distinguish between strain —transferred through the thermal compression of PC—and merely thermally induced effects.

Two-terminal resistance measurements as a function of temperature were simultaneously performed for CrSBr devices prepared on $SiO_2$/Si and PC (Figure 1c). At room-temperature, CrSBr flakes show resistance values in the range of a few megaohms (MΩ) on both substrates. As a function of temperature, CrSBr conductance (*G*) displays variations linked to its semiconducting nature and to magnetic phase transitions. Bulk CrSBr undergoes a magnetic phase transition from paramagnetic to antiferromagnetic order at the Néel temperature $T_N$~132 K [36,38]. In few-layer flakes, this transition temperature remains nearly unchanged in few-layer flakes[9,45], and has been identified by a local maximum in conductance in previous studies[9,36,45]. In Figure 1c, we can observe a slight shift in the maximum of *G* towards a lower value of temperature for CrSBr flakes on PC compared to $SiO_2$/Si (the derivative of conductance as a function of temperature is shown in Figure S1). While our data cannot provide an exact determination of $T_N$, the observed shift suggests a reduction of ~10 K for CrSBr samples on PC. We attribute this behavior to the induced compressive strain in CrSBr due to the thermal contraction of the substrate. During cooldown from 300 to 10 K, polycarbonate substrates have been shown to contract by ~1.2%, with an expected compression of ~0.77% near the $T_N$ of CrSBr [33]. A more detailed discussion on the impact of strain-induced changes in critical temperatures is presented later in the manuscript, supported by magnetoresistance data and first principles calculations.

Then, we performed variable temperature Raman spectroscopy of CrSBr flakes deposited on $SiO_2$/Si and PC substrates (Figure 1d). Raman spectra for CrSBr exhibit three out-of-plane vibration modes, namely $A_g^1$, $A_g^2$, and $A_g^3$ [43,46]. Our analysis focuses on the thermal evolution of $A_g^2$ and $A_g^3$ (Figure 1d and S2). For CrSBr on $SiO_2$/Si, both $A_g^2$ and $A_g^3$ modes exhibit similar blueshifts (by ~5 $cm^{-1}$ and 3 $cm^{-1}$, respectively) upon cooling from 293 to 30 K, following the expected thermal evolution[47]. By contrast, in the case of CrSBr on PC, the $A_g^2$ mode exhibits a frequency shift of ~8 $cm^{-1}$ and the $A_g^3$ of ~9 $cm^{-1}$. This results in a remarkable variation of 6 $cm^{-1}$ for $A_g^3$ and a more subtle change of 3 $cm^{-1}$ for $A_g^2$, which can be attributed to the strain-induced effects of the PC substrate (Figure 1e). According to Ref. [43], Br atoms, which mediate the interlayer coupling, are less involved in the $A_g^3$ mode than in the $A_g^2$ mode, giving $A_g^3$ a weaker interlayer character. Modes with a more marked intralayer character are expected to be more sensitive to the effects of biaxial compressive strain, explaining the additional shift observed for the $A_g^3$ mode in PC substrates compared to $SiO_2$/Si. To further validate this observation, we conducted phonon calculations for CrSBr at different strain values (Figure S4 and S5), confirming that CrSBr remains stable and retains its Pmmn crystal symmetry. For the pristine material we obtained Raman frequency modes of 123 $cm^{-1}$, 248 $cm^{-1}$, and 329 $cm^{-1}$ for $A_g^1$, $A_g^2$, and $A_g^3$, respectively. Our calculations demonstrate that both $A_g^2$ and $A_g^3$ shift to higher frequencies under compressive strain (Figure 1f), with $A_g^3$ exhibiting a significantly larger variation —approximately twice than that of $A_g^2$—, corroborating the strain origin of the observed experimental shifts (Figure 1e).

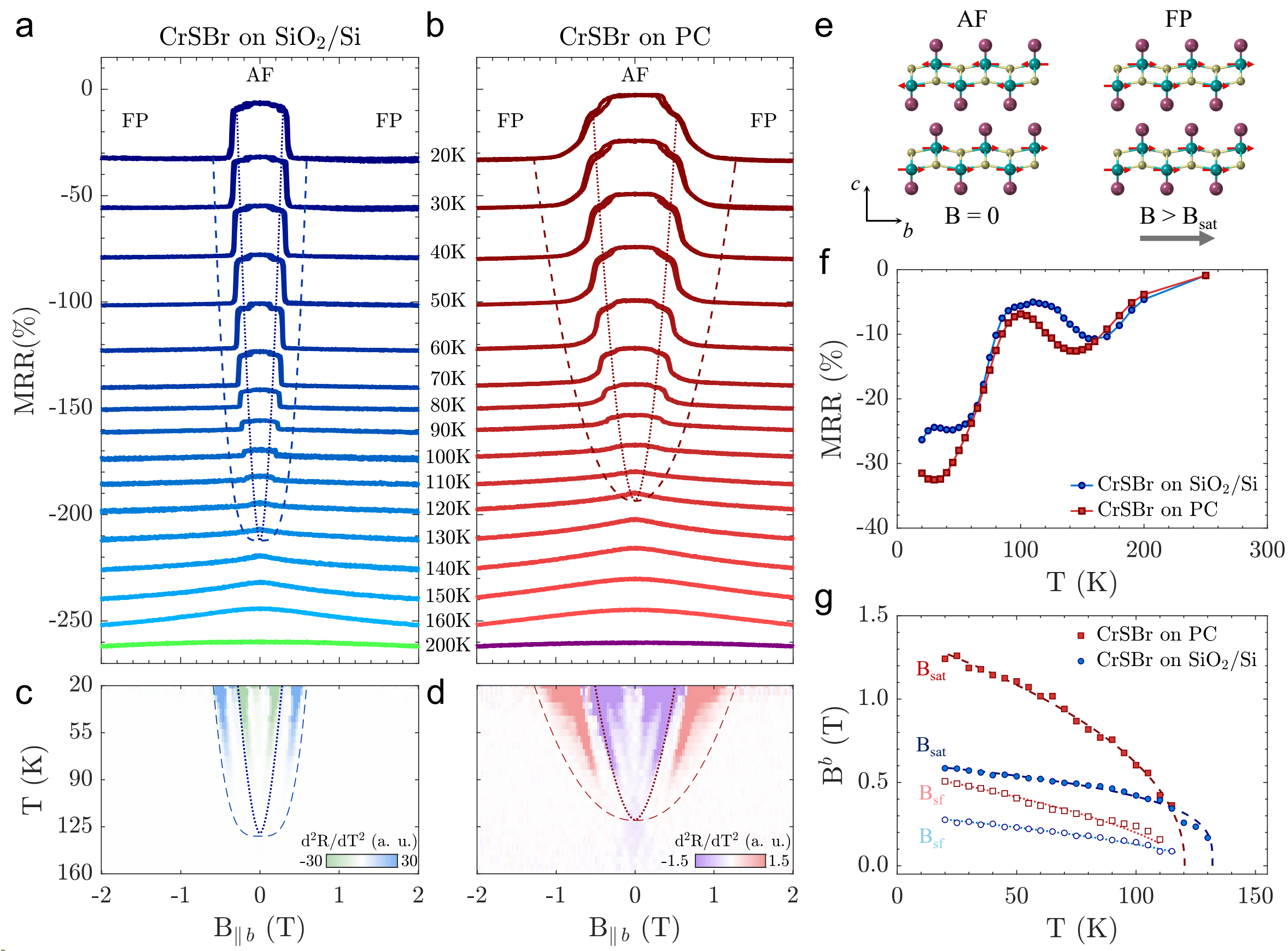

**Figure 2. Magnetoresistance for magnetic field applied along the *b*-axis. a-b)** Magnetoresistance ratio (MRR) for CrSBr flakes on $SiO_2$/Si (a) and PC substrates (b), as a function of temperature and applied magnetic field parallel to the magnetic easy axis (*b*-axis) of CrSBr. Curves have been vertically offset for clarity. Dashed lines indicate the evolution of saturation fields ($B_{sat}$) with temperature, separating the low-field antiferromagnetic (AF) state from the fully-polarized (FP) state. Dotted lines indicate the evolution of the spin-flop transition critical field ($B_{sf}$). **c-d)** Second derivative of resistance versus temperature $d^2R/dT^2$ as a function of applied magnetic field for CrSBr flakes on $SiO_2$/Si (c) and PC substrates (d). Dashed and dotted lines respectively represent the evolution of the saturation and spin-flop fields with temperature. **e)** Crystal structure of CrSBr, with red arrows indicating magnetization orientation for (left) the AF phase, in the absence of an external magnetic field; (right) the FP phase for magnetic field $B>B_{sat}$ applied along the crystallographic *b*-axis. **f)** MRR calculated at B = 3 T as a function of temperature for CrSBr on $SiO_2$/Si and PC substrates. **g)** Saturation field values as a function of temperature, extracted from data in panels a-d. Dashed and dotted lines represent a fit of saturation and spin-flop field data to a Curie-Weiss-like power law model (see main text).

After confirming the induction of biaxial compressive strain for CrSBr on PC upon cooling, we investigate its effects on the magnetoresistance of the devices (Figure 2). Magnetotransport measurements were simultaneously performed on both CrSBr devices on $SiO_2$/Si and PC (see Methods section). First, we characterize the magnetoresistance as a function of temperature for applied magnetic fields along *b* ($B_{||b}$), i.e. parallel to the easy axis of magnetization (Figure 2a-d). At 20 K and zero field, CrSBr is in the antiferromagnetic (AF) state (illustrated in Figure 2e). As the applied field increases, the magnetoresistance ratio (MRR), defined as $MRR = \frac{R(B)-R\,(B=0)}{R(B=0)} \times 100$ exhibits a clear step-like drop (Figure 2a,b). This behavior has been attributed to the alignment of spins with the applied magnetic field, switching from the AF state to a fully polarized (FP) state (sketched in Figure 2e). This variation in the MRR arises from the influence of magnetic order on interlayer tunneling, which is suppressed in the AF phase and restored in the FP state[9]. As the magnetic field increases further, the MRR gradually saturates to a constant value. The critical field at which this saturation occurs defines the saturation field ($B_{sat}$), determined as the field value where $d^2R/dB^2$ approaches zero (Figure 2c,d).

A comparison between panels a and b in Figure 2 reveals striking differences in the MRR characteristics of CrSBr flakes on $SiO_2$/Si and PC substrates. Flakes on PC show broader steps, indicating much higher saturation fields ($B_{sat}$) and exhibit larger changes in the magnitude of their negative MRR (nMRR). Indeed, the nMRR at $B$ = 3 T (Figure 2f) is consistently larger for CrSBr on PC substrates at low temperatures up to 60 K. In particular, below 50 K, the nMRR at 3T for strained CrSBr reaches a maximum intensity of 32% in comparison to 26 % in $SiO_2$/Si (Figure 2f, Table S1). As temperature increases, the magnitude of the nMRR decreases for both samples, reaching a minimum–visible as a peak in Figure 2f– at ~120 K and ~110 K for samples on $SiO_2$/Si and PC, respectively. This difference points to a slight decrease of $T_N$ for CrSBr on PC, since this nMRR minimum has been associated with the antiferromagnetic phase transition[9]. Above this temperature, the magnitude of the nMMR increases again reaching a local maximum at around 170 K for CrSBr on $SiO_2$/Si (appearing as a valley in Figure 2f). This behavior stems from the competition between the remaining intralayer ferromagnetic ordering and the suppression of spin fluctuations by the increasing magnetic field[9]. This temperature, $T^*$ ~ 170 K, linked to the onset of intralayer ferromagnetic correlations, is expected to be independent of the interlayer coupling and to follow the Curie temperature ($T_C$) of individual CrSBr layers[9]. For CrSBr on PC, the nMRR valley shifts to lower temperatures ($T^*$ ~150 K), compared to $SiO_2$/Si, likely due to the biaxial strain induced by the PC substrate compression. In general, strain appears to shift the characteristic features of the MRR curves to lower temperatures, with an overall increase of their magnitude (Figure S6).

Regarding the evolution of saturation magnetic fields in CrSBr flakes with temperature (Figure 2g), we observe noticeable differences between both substrates. Specifically, at 20 K, the saturation field $\mathrm{B}_{\mathrm{sat}}^{b}$ for $SiO_2$/Si, is approximately 0.6T, while this value nearly doubles for CrSBr flakes on PC, reaching $\mathrm{B}_{\mathrm{sat}}^{b}$ ~1.2 T. Moreover, $\mathrm{B}_{\mathrm{sat}}^{b}$ on PC remains consistently higher for temperatures below $T_N$. The observed enhancement of magnetoresistance and $\mathrm{B}_{\mathrm{sat}}^{b}$ in Figure 2 is consistent across different few-layer samples, regardless of the specific thickness and width of the flakes (see Figure S7, S8 and Table S3). A more detailed analysis of the magnetoresistance data reveals the presence of an intermediate step at field values below $\mathrm{B}_{\mathrm{sat}}^{b}$. This has been attributed to a spin-flop transition at critical field $B_{sf}$ [45]. In CrSBr, this transition involves a reorientation of spins from the zero-field AF configuration along the *b* axis to a canted AF state in the *ab* plane above $B_{sf}$ [45]. Notably, $B_{sf}$ is also consistently higher for samples on PC compared to those on $SiO_2$/Si.

From the evolution of $\mathrm{B}_{\mathrm{sat}}^{b}$ with temperature we determined $T_N$ values of ~120 K for PC and ~130 K for $SiO_2$/Si by fitting our data to a modified Curie-Weiss model $T = T_N - (\alpha B_{sat})^{\beta}$ and extrapolating to $\mathrm{B}_{\mathrm{sat}}^{b}$ = 0 T. This was further corroborated by conducting SQUID magnetometry measurements (see Methods and Figure S9) in thin CrSBr films produced by roll-to-roll mechanical exfoliation[48], on both substrates, obtaining $T_N$ values of 118 K and 127 K for the flakes on PC and $SiO_2$/Si, respectively.

To confirm the strain-driven origin of the observed modifications in the magnetic properties of CrSBr upon PC thermal compression, we performed first-principles calculations. We investigated the impact of biaxial strain (ε) on (i) magnetic exchange interactions, (ii) anisotropy energy, (iii) $T_C$ and $T_N$ and (iv) the critical fields ($B_{sat}$ and $B_{sf}$) along the different axes. Note that the magnetic picture for each CrSBr layer can be described by the exchange couplings $J_1$, $J_2$ and $J_3$, which correspond to the interactions between first, second and third neighboring Cr atoms, located along the *a*, diagonal and *b* directions, respectively (see Figure 3a). The computed magnetic exchange parameters for bulk CrSBr are $J_1$ = 0.77 meV, $J_2$ = 3.16 meV and $J_3$ = 7.67 meV, aligning well with prior studies[49–51]. Upon compressive strain up to 1%, the ferromagnetic character of $J_2$ and $J_3$ slightly increases by 0.20 and 0.25 meV, respectively (Figure 3b). On the other hand, $J_1$ varies from 0.77 meV at 0% of strain to -0.67 meV at -1%, resulting in a rapid reduction of $J_1$ by 1.44 meV. Additionally, we evaluated the variation of the interlayer exchange coupling ($J_{int}$), which is defined as the energy difference between ferromagnetic and antiferromagnetic configurations. Therefore, a positive value indicates a more stable AF state. From our calculations, we obtain that $J_{int}$ = 106 μeV/Cr, which agrees well with previous studies (Table S4), thus confirming that bulk CrSBr is an A-type antiferromagnet. Upon applied compressive strain, the AF behavior of CrSBr increases linearly (Figure 3c), reaching a value of $J_{int}$ = 353 μeV/Cr at ε = -1%, indicating an enhancement of the AF character of the system under mechanical deformation. In terms of magnetic anisotropy, we determined values of anisotropy constants of $K_a$ = 46 μeV/Cr and $K_c$ = 137 μeV/Cr for *a* and *c* axes, respectively, correctly capturing the triaxial magnetic anisotropy nature of CrSBr (see Supporting information S9 for details). Notably, similar to $J_{int}$, both $K_a$ and $K_c$ exhibit a linear increase with applied strain (Figures 3d,e). From the results of $J_{int}$, $K_a$ and $K_c$ we computed the spin-flop fields along the *b* axis ($B_{sf}^b$) as well as the saturation fields along the *b*, *a* and *c* axes, namely $B_{sat}^b$, $B_{sat}^a$ and $B_{sat}^c$, respectively (Figures 3f,g).

These parameters were computed using a model that considers two CrSBr sublattices coupled antiferromagnetically. The transition between the antiferromagnetic and ferromagnetic states is described using the following expression for the free energy:

$$F = \frac{J_{int}}{2}\boldsymbol{m}_1\boldsymbol{m}_2 + \frac{K_a}{2}((m_1^a)^2 + (m_2^a)^2) + \frac{K_c}{2}((m_1^c)^2 + (m_2^c)^2) - M_0\boldsymbol{B}\frac{\boldsymbol{m}_1+\boldsymbol{m}_2}{2} \quad (1)$$

This expression assumes that the ferromagnetic order within CrSBr layers is always preserved, while the magnetization of the layers can rotate under an external magnetic field. This behavior arises from the stronger character of the intralayer exchange interactions with respect to Zeeman energy. However, the latter is comparable to both interlayer exchange and the anisotropy energy. In Eq. (1), *F* represents the magnetic free energy per Cr atom, $\boldsymbol{m}_1$ and $\boldsymbol{m}_2$ are the directions of magnetizations of the first and second sublattices. Additionally, $m_1^a$ and $m_1^c$ correspond to the projections of the direction $\boldsymbol{m}_1$ on the *a*- and *c*-axis respectively and $M_0 = \left(\frac{3}{2}\right) g\mu_b$ is the magnetic moment of a single Cr atom.

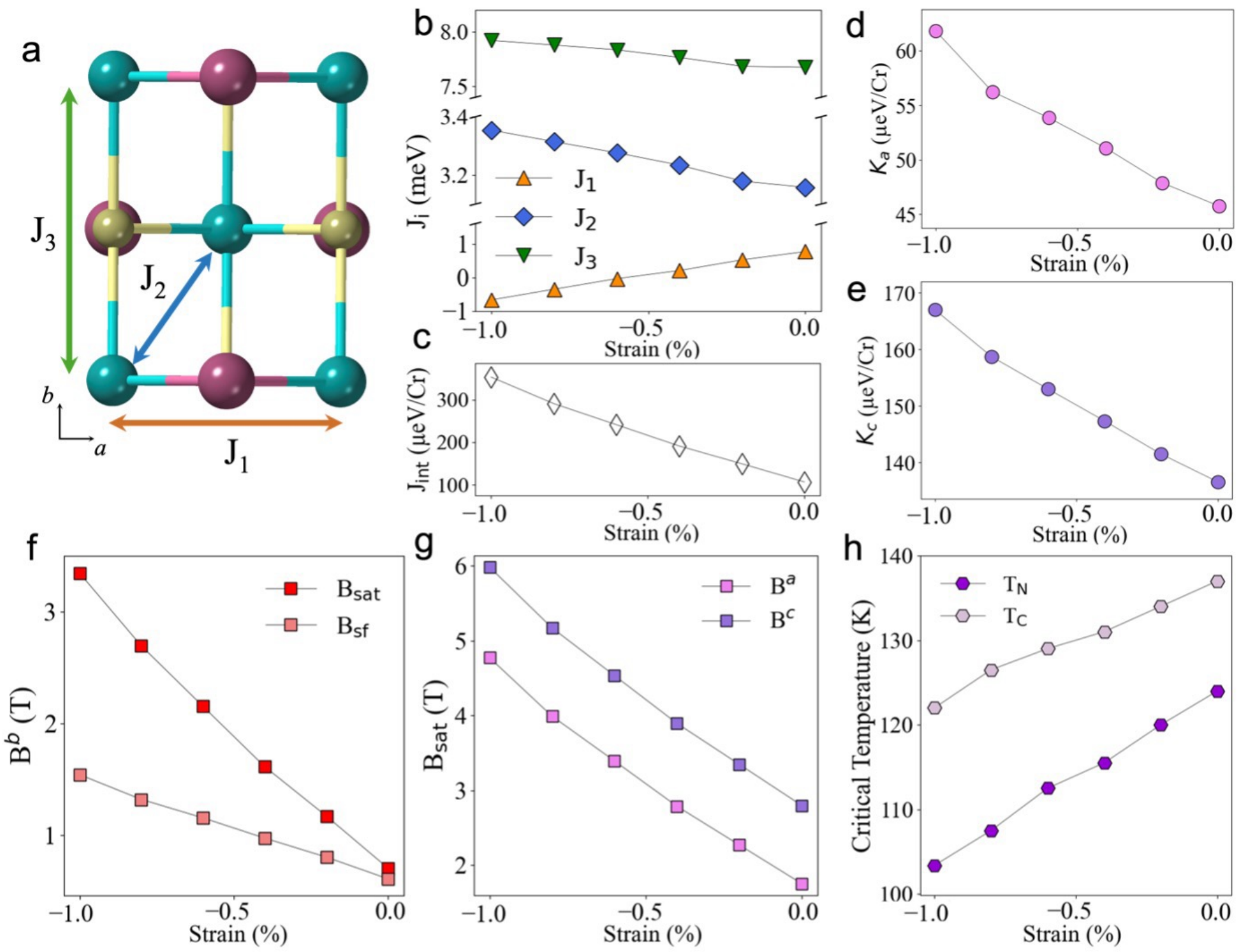

**Figure 3. First-principles calculations of exchange couplings, magnetic anisotropy, critical temperatures and saturation fields.** a) Top view of the crystal structure of a single CrSBr layer, illustrating the magnetic exchange interactions $J_1$, $J_2$ and $J_3$ for the first, second, and third nearest neighbors, represented by red, blue and green arrows, respectively. Evolution of b) intralayer $J_1$, $J_2$ and $J_3$; c) interlayer $J_{int}$ coupling; d) $K_a$; e) $K_c$; f) $B^b_{sat}$ and $B^b_{sf}$; g) $B^a_{sat}$ and $B^c_{sat}$; and h) $T_N$ and $T_C$ as a function of biaxial compressive strain.

When an external magnetic field is applied along the *b*-axis, the spins in each CrSBr layer undergo a sudden rotation from a collinear AF configuration along the *b*-axis to a state where they are tilted within the *ab* plane. The critical field at which this transition initially occurs is referred to as the spin-flop field ($B^b_{sf}$) [52]. For fields exceeding $B^b_{sf}$, spins experience a continuous rotation until reaching the saturation field ($B^b_{sat}$), above which they remain polarized (FP) along *b*. Conversely, when the field is applied along the intermediate *a*-axis or the hard *c*-axis, there is continuous tilting of spins until reaching a FP configuration at the respective saturation fields $B^a_{sat}$ and $B^c_{sat}$ (see Supporting Information S10 for a more detailed discussion). The expressions used to determine these fields are given by:

$$B^b_{sf} = \frac{2\sqrt{K_a(J_{int} - K_a)}}{M_0} \qquad (2)$$

$$B^b_{sat} = 2\frac{J_{int} - K_a}{M_0} \qquad (3)$$

$$B_{sat}^{a} = 2\frac{J_{int} + K_a}{M_0} \quad (4)$$

$$B_{sat}^{c} = 2\frac{J_{int} + K_c}{M_0} \quad (5)$$

For applied fields along the *b*-axis, both predicted $B_{sat}^{b}$ and $B_{sf}^{b}$ increase substantially with biaxial compressive strain (Figure 3f). Although $K_a$ increases upon compressive strain and its contribution is subtracted in Eqs. (2) and (3), the primary factor governing the behavior of $B_{sat}^{b}$ and $B_{sf}^{b}$ is $J_{int}$. This is because $J_{int}$ not only possesses a larger magnitude but also exhibits a more pronounced increase under compression compared to $K_a$. This analysis confirms that compressive strain reinforces the antiferromagnetic interlayer coupling in CrSBr, effectively shifting the critical fields for spin-flop and spin-flip transitions to higher values, in agreement with experimental trends.

Specifically, at 0% of strain, we obtained that $B_{sat}^{b}$= 0.7T and $B_{sf}^{b}$ = 0.6T, which are enhanced up to values of $B_{sat}^{b}$ = 3.35T and $B_{sf}^{b}$ = 1.54T at ε = -1%. This trend agrees well with experimental findings (Figure 2g), which show that both $B_{sat}^{b}$ and $B_{sf}^{b}$ are higher in CrSBr on PC, confirming their origin in the induced compressive strain. Experimentally, $B_{sat}^{b}$ increases from 0.6 to 1.2 T for CrSBr on $SiO_2$/Si and PC substrates at 20K, respectively, yielding a twofold enhancement ($\frac{B_{sat}^{b}(PC)}{B_{sat}^{b}(SiO_2/Si)} \sim 2$). A similar ratio is obtained from our calculations $\frac{B_{sat}^{b}(\varepsilon)}{B_{sat}^{b}(0)} \sim 2$ for a compressive strain of ε ~0.33%. This theoretical estimation provides a lower bound for the transferred strain from the thermal compression of polycarbonate —which offers an ideal upper bound of ~1% at 20 K—since calculations do not account for finite temperature effects that can reduce the experimentally observed saturation fields. Furthermore, our calculations predict similar enhancements in the saturation fields along the intermediate *a*- and the hard *c*-axis (Figure 3g). This aspect will be further discussed when presenting the experimental results of magnetoresistance for magnetic fields applied along the intermediate and hard directions (Figure 4).

Figure 3h illustrates the evolution of $T_N$ under mechanical deformation, revealing a linear decrease which is primarily driven by the rapid reduction of the ferromagnetic contribution of $J_1$. Consistent with experimental observations (Figure 2f), a reduction of 5-10K of $T_N$ is expected for the reported strain level. In addition, we computed the evolution of $T_C$ for a CrSBr single layer, which exhibits a similar strain-induced decrease, aligning closely with the experimentally observed shift of *T** towards lower temperatures for CrSBr on PC compared to $SiO_2$/Si. Since monolayers lack interlayer coupling, the $T_C$ evolution purely reflects changes in intralayer exchange. The observed similarity between the strain dependence of $T_C$ in monolayer and $T_N$ in bulk CrSBr proves that the primary driving force behind the variation of $T_N$ is the modulation of intralayer interactions and thus it is also a fundamentally two-dimensional effect.

Magnetoresistance data for magnetic fields applied along the intermediate *a-axis* ($B_{||a}$) and hard *c*-axis ($B_{||c}$) are shown in Figure 4a-b and f-g, respectively. In both cases, samples

on PC exhibit a clear increase in the intensity and width of the MRR domes compared to those on $SiO_2$/Si, consistent with the behavior observed for $B_{||b}$. Specifically, for CrSBr on PC, the nMRR is enhanced by approximately 7% at 3 T for fields applied along the a-axis, and by 8% at 5 T for fields applied along the c-axis. Regarding the evolution of $B_{sat}$, both $B_{sat}^{a}$ and $B_{sat}^{c}$ are enhanced when CrSBr is on PC compared to $SiO_2$/Si (Figure 4e,j). This behavior is predicted by our first-principles calculations (Figure 3), which show that compressive strain increases the interlayer exchange coupling ($J_{int}$) and as well as the magnetic anisotropy energies $K_a$ and $K_c$. These changes lead to corresponding increases in $B_{sat}^{a}$ and $B_{sat}^{c}$, as described by Eqs. (4) and (5), respectively.

At 20K, when the field is applied along the *a*-axis, $B_{sat}^{a}$ increases from ~1.3 T on $SiO_2$/Si to ~2 T on PC due to the induced strain, yielding a ratio of $B_{sat}^{a}(PC)/B_{sat}^{a}(SiO_2/Si)$ ~ 1.6. Similarly, for fields along the *c*-axis, $B_{sat}^{c}$ is enhanced by a factor of ~ 1.4 times. These findings are in excellent agreement with our theoretical predictions which estimate enhancement ratios of ~1.5 and ~1.3, for the *a*- and *c*-axis respectively under a strain level of 0.33% (see Figure 3g and Table S2). Finally, the analysis of the temperature evolution of $B_{sat}$ (Figure 4e,j) along the *a*- and *c*-axis yields $T_N$ ~ 120 K and 130 K for PC and $SiO_2$/Si substrates. These results are consistent with those obtained for $B_{||b}$ and our theoretical calculations, confirming the subtle reduction of $T_N$ due to compressive strain, primarily driven by a weakening of the first-neighbor intralayer exchange ($J_1$).

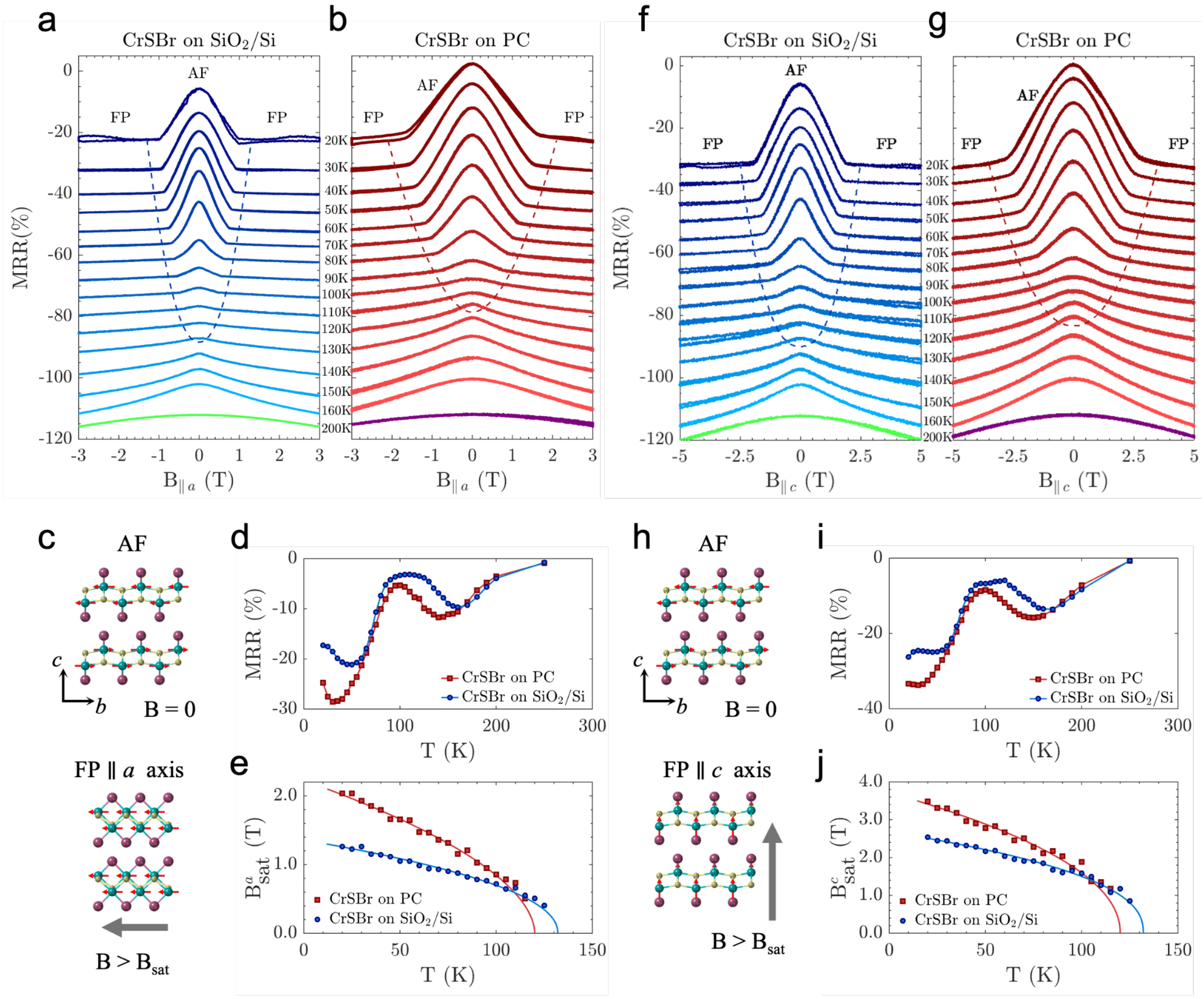

**Figure 4. Magnetoresistance for magnetic field applied along the *a*- (a-e) and *c*-axis (f-j). a-b)** Magnetoresistance ratio (MRR), for CrSBr flakes on a) $SiO_2$/Si and b) PC substrates with $B_{||a}$. **c)** Crystal structure of CrSBr, with red arrows indicating magnetization orientation for (top) the AF phase; (bottom) the FP phase for $B > B^a_{sat}$. d) MRR at B = 3 T as a function of temperature, and **e)** saturation field values as a function of temperature, for CrSBr on $SiO_2$/Si and PC substrates and $B_{||a}$. **f-g)** Magnetoresistance ratio (MRR), for CrSBr flakes on f) $SiO_2$/Si and g) PC substrates for $B_{||c}$. **h)** Crystal structure of CrSBr, with red arrows indicating magnetization for (top) the AF phase; (bottom) the FP phase for $B > B^c_{sat}$. **i)** MRR at B = 5 T as a function of temperature and **j)** saturation field values as a function of temperature, for CrSBr on $SiO_2$/Si and PC substrates and $B_{||c}$.

## Conclusions

In summary, we demonstrate effective strain-tuning of the magnetic properties in the 2D layered antiferromagnet CrSBr, as revealed by magnetotransport measurements and comparison with first-principles calculations. Specifically, compressive strain significantly increases device magnetoresistance and saturation fields for applied fields along all three crystallographic directions, with a two-fold enhancement in the field required to polarize the material along its easy magnetic axis. These effects are explained by a dramatic enhancement of the interlayer exchange coupling and magnetic anisotropy energy. Additionally, compressive strain leads to a modest 10K reduction in both $T_N$ and $T_C$, driven by a weakening of the ferromagnetic character of the first-neighbor intralayer exchange interactions.

Our findings deepen the understanding of how strain modifies the magnetotransport properties of CrSBr, targeting the unexplored regime of biaxial compressive strain. This work complements and extends recent studies focused on uniaxial tensile strain. We demonstrate that strain engineering can be a robust and cost-effective approach for tailoring 2D magnetism, using thermally induced strain from substrate mismatch. By directly linking experimental magnetotransport responses to strain-induced variations in exchange couplings and magnetic anisotropy from first-principles modeling, we provide a comprehensive picture of the effects of biaxial compressive strain in CrSBr. Moreover, the demonstrated ability to enhance and control magnetic anisotropy and magnetoresistance offers a practical route toward the design of strain-programmable 2D magnetic and spintronic devices.

## Materials and Methods

**Sample preparation.** A CrSBr crystal (HQ Graphene) was mechanically exfoliated using 1008 blue silicon-free tape (Medium-High Tack, Ultron Systems Inc.) onto a transparent polydimethylsiloxane (PDMS)-based gel film (WF4 x 6.0 mil from Gel-Pak) for inspection under a modified optical metallurgical microscope. The flakes thicknesses were initially estimated using transmittance spectroscopy[53] while they were still on the PDMS substrates. Pairs of flakes with similar thickness were selected, with each flake in the pair transferred, via a dry transfer method, onto either a 250 µm thick polycarbonate film or a Si/$SiO_2$ substrate (300 nm oxide layer), both with pre-patterned Ti/Au (5 nm/45 nm) electrodes and 30 µm channel length fabricated by electron beam evaporation using a metal shadow mask from Ossila Ltd. The thickness values obtained from transmittance spectroscopy prior to transfer were later confirmed by atomic force microscopy (AFM) after the flakes were transferred onto $SiO_2$/Si or PC substrates and all other measurements had been completed. Thickness values reported in the main text and Supporting Information correspond to those obtained from AFM measurements.

**Magnetotransport measurements.** Pairs of CrSBr devices of similar thickness fabricated on PC and on SiO2/Si substrates were loaded together and simultaneously measured in an Oxford Instruments Teslatron VTI and in a MPMS Quantum Design cryostat. In both cases, temperature was swept at ~1 K/min and resistance was measured in a two-terminal constant voltage ($V_{AC}$~1mV, at 17.777 Hz) configuration using SRS830 lock-in amplifiers. Magnetic field sweeps were performed at a rate of 0.27 T/min, ranging from -5T to 5T along the c-axis and from -3T to 3T along both the *a*-axis and *b*-axis. In the case of $SiO_2$/Si devices, the silicon gate was kept grounded throughout the entire measurement process.

**Biaxial compressive strain engineering**. Biaxial compressive strain was achieved by exploiting the thermal mismatch between CrSBr flakes and a polycarbonate (PC) substrate upon cooling, the larger thermal contraction of PC (~1.2%) induces compressive strain in the CrSBr flakes. This method has been demonstrated to successfully induce biaxial compressive strain to 2D materials (Refs. [32–34,44]). The cooldown and warmup cycles do not appear to damage the flakes (Figure S15).

**Variable temperature Raman spectroscopy.** Raman spectroscopy was performed for flakes of similar thickness (~ 10 nm) to those used in the transport measurements deposited by similar means on $SiO_2$/Si and PC substrates, respectively. Temperature dependent Raman spectra were recorded using unpolarized excitation with a wavelength of 514 nm. A McPherson 207 monochromator equipped with an Andor newton EMCCD detector were used for analysis of light reflected by the samples mounted in a Montana Instruments optical cryostat. Each spectrum was acquired with 2 acquisitions of 90 s and a grid of 1200 was used. A linear background was subtracted from the Raman spectra collected on PC, likely arising from broad emission of the material, particularly strong at low temperatures (see Figure S3).

**First-principles calculations.** We performed spin polarized density-functional theory (DFT) calculations on bulk CrSBr using the VASP package[54]. The generalized gradient approximation (GGA) was employed as the exchange-correlation functional. The nonlocal functional optB86b was used to account for the van der Waals (vdW) interactions between layers, as it has been proved to properly capture the properties of CrSBr and other magnetic layered materials[55–57]. The antiferromagnetism between adjacent CrSBr layers was considered by using a supercell of dimensions 1 × 1 × 2. The calculations for the monolayer were performed by adding an 18 Å vacuum space in the perpendicular direction to avoid undesirable interactions between adjacent layers (see Supporting Information S11 for details). The Brillouin zone was sampled by well converged Γ-centered 20×20×2 (20×20×1) k-point Monkhorst−Pack mesh for bulk (monolayer) calculations. We did not include Hubbard U correction to treat localized d electrons given as it has been proved to wrongly capture the antiferromagnetism in CrSBr [58,59]. The Raman frequency modes were determined using the Phonopy package and employing supercells of dimensions 3 × 3 × 2 [60]. To determine the intralayer exchange couplings $J_1$, $J_2$ and $J_3$ we employed the TB2J code[61], using supercells of dimensions 20 × 20 × 1. As an input for the TB2J code, we used a tight binding Hamiltonian on the basis on maximally localized Wannier functions (MLWFs). It was constructed by means of the Wannier90 package[62] by using the *d* orbitals of Cr atoms and the *p* orbitals of both Br and S as the basis set. The values of $T_N$ and $T_C$ of CrSBr were obtained by performing atomistic simulations by means of the VAMPIRE code[63]. We selected supercells of dimensions 20 nm × 20 nm × 20 nm, with both equilibration and averaging phases performed using 10,000 steps.

**Magnetometry in CrSBr films.** SQUID magnetometry measurements were performed on two sets of samples prepared on different substrates: polycarbonate and $SiO_2$/Si. Each sample was coated with a thin film (10–50 nm thick) comprising a densely packed network of CrSBr flakes produced by *roll-to-roll* mechanical exfoliation, achieving a surface coverage greater than 80%. Magnetization measurements were performed in a Quantum Design MPMS-5S SQUID magnetometer mounted in a low-temperature cryostat, under a 1000 Oe field approximately aligned along the *a*-axis of CrSBr flakes. Each sample was secured inside a diamagnetic capsule.

## Acknowledgements

The authors acknowledge funding from the Spanish government through grants RED2022-134448-T, PID2023-146354NB-C41, PID2023-146354NB-C44, PID2022-136285NB-C32, PDC2023-145920-I00, PID2023-151946OB-I00, TED2021-132267B-I00, PID2020-115566RB-I00 (all funded by MICIU/AEI /10.13039/501100011033, and from EU FEDER), CNS2023-145151 (funded by MICIU/AEI/10.13039/501100011033 and from EU NextGenerationEU/PRTR), the RyC Fellowships (RYC2018-024736-I to E.N.M. and RYC2021-034609-I to M.G.), and the Spanish Unidad de Excelencia "María de Maeztu" (CEX2019-000919-M). J.J.B. acknowledges the European Union (ERC-2021-StG-101042680 2D-SMARTiES) and the Generalitat Valenciana (grant CIDEXG/2023/1). E.N.M. acknowledges the European Research Council (ERC) under Horizon 2020 research and innovation program (ERC StG, grant agreement No. 803092). A.C.G. acknowledges funding from the European Union through grant ERC-2024-PoC StEnSo (grant agreement 101185235). M.G. thanks the Generalitat Valenciana for the GenT grant CISEJI/2023/45. A.M.R. thanks the Spanish MIU (Grant No FPU21/04195). E.D., A.P.R., and M.A. acknowledge support from FEDER/Junta de Castilla y León Research (Grant SA106P23). J.S.S. acknowledges financial support from the Consejería de Educación, Junta de Castilla y León, and ERDF/FEDER. A.P.R. acknowledges the financial support received from the Marie Skłodowska Curie-COFUND program under the Horizon 2020 research and innovation initiative of the European Union, within the framework of the USAL4Excellence program (Grant 101034371).

## References

1. Lee, J.-U. *et al.* Ising-Type Magnetic Ordering in Atomically Thin FePS3. *Nano Lett.* **16**, 7433–7438 (2016).
2. Huang, B. *et al.* Layer-dependent ferromagnetism in a van der Waals crystal down to the monolayer limit. *Nature* **546**, 270–273 (2017).
3. Gong, C. *et al.* Discovery of intrinsic ferromagnetism in two-dimensional van der Waals crystals. *Nature* **546**, 265–269 (2017).
4. Gibertini, M., Koperski, M., Morpurgo, A. F. & Novoselov, K. S. Magnetic 2D materials and heterostructures. *Nat. Nanotechnol.* **14**, 408–419 (2019).
5. Wu, F. *et al.* Quasi-1D Electronic Transport in a 2D Magnetic Semiconductor. *Adv. Mater.* **34**, 2109759 (2022).
6. Wang, F. *et al.* New Frontiers on van der Waals Layered Metal Phosphorous Trichalcogenides. *Adv. Funct. Mater.* **28**, 1802151 (2018).
7. Fu, Z. *et al.* Anomalous Tunneling Magnetoresistance Oscillation and Electrically Tunable Tunneling Anisotropic Magnetoresistance in Few-Layer CrPS4. *Adv. Phys. Res.* **3**, 2400052 (2024).
8. Grubišić-Čabo, A. *et al.* 'Roadmap on quantum magnetic materials', Antonija Grubišić-Čabo et al 2025 2D Mater. 12 031501 https://doi.org/10.1088/2053-1583/adbe89
9. Telford, E. J. *et al.* Coupling between magnetic order and charge transport in a two-dimensional magnetic semiconductor. *Nat. Mater.* **21**, 754–760 (2022).
10. Wang, Z. *et al.* Very large tunneling magnetoresistance in layered magnetic semiconductor CrI3. *Nat. Commun.* **9**, 2516 (2018).

11. Jo, J. *et al.* Nonvolatile Electric Control of Antiferromagnet CrSBr. *Nano Lett.* **24**, 4471–4477 (2024).
12. Chen, Y. *et al.* Twist-assisted all-antiferromagnetic tunnel junction in the atomic limit. *Nature* **632**, 1045–1051 (2024).
13. Park, S. Y. *et al.* Controlling the Magnetic Anisotropy of the van der Waals Ferromagnet Fe3GeTe2 through Hole Doping. *Nano Lett.* **20**, 95–100 (2020).
14. Qi, Y. *et al.* Recent Progress in Strain Engineering on Van der Waals 2D Materials: Tunable Electrical, Electrochemical, Magnetic, and Optical Properties. *Adv. Mater.* **35**, 2205714 (2023).
15. Roldán, R., Castellanos-Gomez, A., Cappelluti, E. & Guinea, F. Strain engineering in semiconducting two-dimensional crystals. *J. Phys. Condens. Matter* **27**, 313201 (2015).
16. Xie, W.-Q., He, C.-C., Yang, X.-B., Zhao, Y.-J. & Geng, W.-T. Strain-Induced Interlayer Magnetic Coupling Spike of the Two-Dimensional van der Waals Material Fe5GeTe2. *J. Phys. Chem. C* **127**, 17194–17200 (2023).
17. Hu, X. *et al.* Enhanced Ferromagnetism and Tunable Magnetism in Fe3GeTe2 Monolayer by Strain Engineering. *ACS Appl. Mater. Interfaces* **12**, 26367–26373 (2020).
18. Mahajan, A. & Bhowmick, S. Decoupled strain response of ferroic properties in a multiferroic VOCl2 monolayer. *Phys. Rev. B* **103**, 075436 (2021).
19. Pizzochero, M. & Yazyev, O. V. Inducing Magnetic Phase Transitions in Monolayer CrI3 via Lattice Deformations. *J. Phys. Chem. C* **124**, 7585–7590 (2020).
20. Webster, L. & Yan, J.-A. Strain-tunable magnetic anisotropy in monolayer CrCl3, CrBr3, and CrI3. *Phys. Rev. B* **98**, 144411 (2018).
21. Lv, H. Y., Lu, W. J., Luo, X., Zhu, X. B. & Sun, Y. P. Strain- and carrier-tunable magnetic properties of a two-dimensional intrinsically ferromagnetic semiconductor: CoBr2 monolayer. *Phys. Rev. B* **99**, 134416 (2019).
22. Yang, K., Wang, G., Liu, L., Lu, D. & Wu, H. Triaxial magnetic anisotropy in the two-dimensional ferromagnetic semiconductor CrSBr. *Phys. Rev. B* **104**, 144416 (2021).

23. Xu, B. *et al.* Switching of the magnetic anisotropy via strain in two dimensional multiferroic materials: CrSX (X = Cl, Br, I). *Appl. Phys. Lett.* **116**, 052403 (2020).

24. Diao, Y. *et al.* Strain-regulated magnetic phase transition and perpendicular magnetic anisotropy in CrSBr monolayer. *Phys. E Low-Dimens. Syst. Nanostructures* **147**, 115590 (2023).

25. Lei, B. *et al.* Modulation of magnetic phase transition and magnetic anisotropy in CrSe2 monolayer under biaxial strain. *J. Appl. Phys.* **137**, 093904 (2025).

26. Esteras, D. L., Rybakov, A., Ruiz, A. M. & Baldoví, J. J. Magnon Straintronics in the 2D van der Waals Ferromagnet CrSBr from First-Principles. *Nano Lett.* **22**, 8771–8778 (2022).

27. Cenker, J. *et al.* Reversible strain-induced magnetic phase transition in a van der Waals magnet. *Nat. Nanotechnol.* **17**, 256–261 (2022).

28. Cenker, J. *et al.* Engineering Robust Strain Transmission in van der Waals Heterostructure Devices. *Nano Lett.* (2025).

29. Cenker, J. *et al.* Strain-programmable van der Waals magnetic tunnel junctions. *Newton* 100130 (2025)

30. Bagani, K. *et al.* Imaging Strain-Controlled Magnetic Reversal in Thin CrSBr. *Nano Lett.* **24**, 13068–13074 (2024).

31. Šiškins, M. *et al.* Magnetic and electronic phase transitions probed by nanomechanical resonators. *Nat. Commun.* **11**, 2698 (2020).

32. Frisenda, R. *et al.* Biaxial strain tuning of the optical properties of single-layer transition metal dichalcogenides. *Npj 2D Mater. Appl.* **1**, 10 (2017).

33. Henríquez-Guerra, E. *et al.* Large Biaxial Compressive Strain Tuning of Neutral and Charged Excitons in Single-Layer Transition Metal Dichalcogenides. *ACS Appl. Mater. Interfaces* **15**, 57369–57378 (2023).

34. Henríquez-Guerra, E. *et al.* Modulation of the Superconducting Phase Transition in Multilayer 2H-NbSe2 Induced by Uniform Biaxial Compressive Strain. *Nano Lett.* **24**, 10504–10509 (2024).

35. Ziebel, M. E. *et al.* CrSBr: An Air-Stable, Two-Dimensional Magnetic Semiconductor. *Nano Lett.* **24**, 4319–4329 (2024).
36. Telford, E. J. *et al.* Layered Antiferromagnetism Induces Large Negative Magnetoresistance in the van der Waals Semiconductor CrSBr. *Adv. Mater.* **32**, 2003240 (2020).
37. Klein, J. *et al.* The Bulk van der Waals Layered Magnet CrSBr is a Quasi-1D Material. *ACS Nano* **17**, 5316–5328 (2023).
38. Göser, O., Paul, W. & Kahle, H. G. Magnetic properties of CrSBr. *J. Magn. Magn. Mater.* **92**, 129–136 (1990).
39. Wilson, N. P. *et al.* Interlayer electronic coupling on demand in a 2D magnetic semiconductor. *Nat. Mater.* **20**, 1657–1662 (2021).
40. Marques-Moros, F., Boix-Constant, C., Mañas-Valero, S., Canet-Ferrer, J. & Coronado, E. Interplay between Optical Emission and Magnetism in the van der Waals Magnetic Semiconductor CrSBr in the Two-Dimensional Limit. *ACS Nano* **17**, 13224–13231 (2023).
41. Krelle, L. *et al.* Magnetic Correlation Spectroscopy in CrSBr. Preprint at https://doi.org/10.48550/arXiv.2503.08390 (2025).
42. Boix-Constant, C. *et al.* Probing the Spin Dimensionality in Single-Layer CrSBr Van Der Waals Heterostructures by Magneto-Transport Measurements. *Adv. Mater.* **34**, 2204940 (2022).
43. Torres, K. *et al.* Probing Defects and Spin-Phonon Coupling in CrSBr via Resonant Raman Scattering. *Adv. Funct. Mater.* **33**, 2211366 (2023).
44. Gant, P. *et al.* A strain tunable single-layer MoS2 photodetector. *Mater. Today* **27**, 8–13 (2019).
45. Ye, C. *et al.* Layer-Dependent Interlayer Antiferromagnetic Spin Reorientation in Air-Stable Semiconductor CrSBr. *ACS Nano* **16**, 11876–11883 (2022).
46. Mondal, P. *et al.* Raman polarization switching in CrSBr. *Npj 2D Mater. Appl.* **9**, 1–7 (2025).

47. Pawbake, A. *et al.* Raman scattering signatures of strong spin-phonon coupling in the bulk magnetic van der Waals material CrSBr. *Phys. Rev. B* **107**, 075421 (2023).

48. Sozen, Y., Riquelme, J. J., Xie, Y., Munuera, C. & Castellanos-Gomez, A. High-Throughput Mechanical Exfoliation for Low-Cost Production of van der Waals Nanosheets. *Small Methods* **7**, 2300326 (2023).

49. Wang, Y., Luo, N., Zeng, J., Tang, L.-M. & Chen, K.-Q. Magnetic anisotropy and electric field induced magnetic phase transition in the van der Waals antiferromagnet CrSBr. *Phys. Rev. B* **108**, 054401 (2023).

50. Telford, E. J. *et al.* Designing Magnetic Properties in CrSBr through Hydrostatic Pressure and Ligand Substitution. *Adv. Phys. Res.* **2**, 2300036 (2023).

51. Liu, J., Zhang, X. & Lu, G. Moiré magnetism and moiré excitons in twisted CrSBr bilayers. *Proc. Natl. Acad. Sci.* **122**, e2413326121 (2025).

52. Coey, J. M. D. *Magnetism and Magnetic Materials*. (Cambridge University Press, 2001). doi:10.1017/CBO9780511845000.

53. Frisenda, R. *et al.* Micro-reflectance and transmittance spectroscopy: a versatile and powerful tool to characterize 2D materials. *J. Phys. Appl. Phys.* **50**, 074002 (2017).

54. Kresse, G. & Furthmüller, J. Efficient iterative schemes for ab initio total-energy calculations using a plane-wave basis set. *Phys. Rev. B* **54**, 11169–11186 (1996).

55. León, A. M., Arnold, B. C., Heine, T. & Brumme, T. Interlayer magnetic coupling in FePS3 and NiPS3 stacked bilayers from first principles. *2D Mater.* **12**, 025023 (2025).

56. Lu, S. *et al.* Controllable dimensionality conversion between 1D and 2D CrCl3 magnetic nanostructures. *Nat. Commun.* **14**, 2465 (2023).

57. Liu, N. *et al.* Intralayer strain tuned interlayer magnetism in bilayer CrSBr. *Phys. Rev. B* **109**, 214422 (2024).

58. Wang, Y., Luo, N., Zeng, J., Tang, L.-M. & Chen, K.-Q. Magnetic anisotropy and electric field induced magnetic phase transition in the van der Waals antiferromagnet CrSBr. *Phys. Rev. B* **108**, 054401 (2023).

59. Xie, K., Zhang, X.-W., Xiao, D. & Cao, T. Engineering Magnetic Phases of Layered Antiferromagnets by Interfacial Charge Transfer. *ACS Nano* **17**, 22684–22690 (2023).

60. Togo, A. & Tanaka, I. First principles phonon calculations in materials science. *Scr. Mater.* **108**, 1–5 (2015).

61. He, X., Helbig, N., Verstraete, M. J. & Bousquet, E. TB2J: A python package for computing magnetic interaction parameters. *Comput. Phys. Commun.* **264**, 107938 (2021).

62. Mostofi, A. A. *et al.* wannier90: A tool for obtaining maximally-localised Wannier functions. *Comput. Phys. Commun.* **178**, 685–699 (2008).

63. Evans, R. F. L. *et al.* Atomistic spin model simulations of magnetic nanomaterials. *J. Phys. Condens. Matter* **26**, 103202 (2014).